\documentclass [aps,prL,amssymb,amsmath,twocolumn,showpacs]{revtex4-1}

\usepackage{microtype}
\pdfoutput=1 
\usepackage{amsmath}
\usepackage{amssymb}
\usepackage{amsthm}
\usepackage{mathrsfs}
\usepackage{graphicx}
\usepackage{verbatim}
\usepackage{bm}
\usepackage[dvipsnames]{xcolor}
\usepackage{graphicx,epsfig,amsfonts,amssymb}
\usepackage{hyperref}
  \hypersetup{colorlinks=true,citecolor=blue}

\usepackage{bm}
\usepackage{times}
\usepackage{lipsum}

\usepackage[all]{xy}

\newcommand{\nn}{\nonumber}

\newcommand{\sgn}{\operatorname{sgn}}
\newcommand{\bk}{\mathbf{k}}

\newcommand{\bh}{\mathbf h}

\newcommand{\ve} {\varepsilon}

\newcommand{\be}{\begin{eqnarray}}
\newcommand{\ee}{\end{eqnarray}}
\newcommand{\la}{\langle}
\newcommand{\ra}{\rangle}
\newcommand{\rar}{\rightarrow}

\begin{document}

\title{Generic theory of characterizing topological phases under quantum slow dynamics}
\author {Panpan Fang$^1$}
\author{Yi-Xiang Wang $^{1,2}$} 
\email{wangyixiang@jiangnan.edu.cn}
\author{Fuxiang Li$^1$}
\email{fuxiangli@hnu.edu.cn}
\affiliation{$^1$School of Physics and Electronics, Hunan University, Changsha 410082, China}
\affiliation{$^2$School of Science, Jiangnan University, Wuxi 214122, China}

\date{\today}

\begin{abstract}
Dynamical characterization of equilibrium topological phases has attracted considerable attention in recent years. In this paper, we make a thorough exploration of  the nonadiabatic characterization of topological phases under slow quench protocol. We first propose an exactly solvable multi-state Landau-Zener model that can be directly applied to the nonadiabatic slow quench dynamics of topological systems. Then we present two different schemes  to characterize the bulk topology of the system based on the so-called spin inversion surface. The first one needs least number of quenching processes, but requires to measure the gradients of time-averaged spin-polarization on the SIS. The second one only needs to measure the value of time-averaged spin-polarization on the SIS, thus making it possible to directly characterize the topological phases by introducing an extra quenching process.  Moreover, high-order SIS or band inversion surface (BIS) relying on the dimension reduction approach, is also generalized to the above two different characterization schemes. One can extract the topological invariant from pairs of points with opposite signs both on the 0D highest order BIS and on the 0D highest order SIS,  which greatly simplifies the measurement strategy and  characterization process. In a word, superior to the sudden quench protocol, we demonstrate that the topological invariant can be captured not only by the topological information on BIS, but also on the SIS. In particular, direct characterization of topological phases based on BIS and SIS can be realized.
 \end{abstract}


\date{\today}

\maketitle


%
\section{Introduction}
In the last two decades, topological quantum phases have ignited extensive attention in both natural and artificial materials \cite{Hasan2010, qi2011,Rechtsman2013, Khanikaev, Ozawa, Miyake2013, Jotzu, Aidelsburger2015, Wu2016}. Compared  with the conventional matter, the most exotic property of the topological quantum phases is that they can support the protected boundary states immune to moderate disorder or defects. In addition, according to the bulk-boundary correspondence (BBC), the number of these boundary states are related to the bulk topological invariant of the topological phases, which, in the equilibrium theory, are defined in the ground state of the system. In consequence, one can identify the topological phases by resolving the boundary modes with angle-resolved photo-electronic spectroscopy (ARPES) and transport measurements experimentally \cite {Hsieh, Xia09, Chang13, Konig, X2019, Xu2015, Lv15}. However, there are still challenges in these experiments to overcome since they are not direct observations of bulk topological numbers \cite{Wilczek2009, Alicea2012, Elliott2015}.

In contrast to the equilibrium theory, there has been increasing interest in using non-equilibrium dynamics as a probe to characterize topological properties of equilibrium topological phases in recent years \cite{Caio, Hu2016, Wang2017, Flaschner, Song2018, McGinley, Lu1911, Qiu2019, Xie2020, Hu2020, Chen2020, Sun2018, Wang2019PRA, Yi2019PRL, Song2019Nat, Ji2002, Xin2001, Niu2001}. Dynamical characterization of topological phases have been studied in various systems, such as Floquet quantum systems, (non-) Hermitian quantum systems, correlated systems, etc \cite{2020winding, nohermitanion, 2019correlate, 2004Floquet, 2021noise}. It is worth noting that a new dynamical characterization theory based on quantum quench dynamics was first proposed theoretically and verified experimentally in Hermitian systems by Liu and his co-workers \cite{2018Sci, experiment1, experiment2, experiment3}. They establish a dynamical bulk-surface correspondence (dBSC), which states a generic $d$D topological phase with ${\mathbb{Z}}$-type invariant can be characterized by the $(d-1)$ D invariant defined on the band inversion surface (BIS). In particular, the dynamical bulk-surface correspondence has its intrinsic advantages in two aspects. First, as a momentum-space counterpart of the BBC in real space, the dBSC tells that the  topological information of the system can be easily measured with high-precision in momentum-space. By suddenly quenching the system from a trivial phase to a topological phase, the bulk topology of the $d$D equilibrium phases of the post-quench Hamiltonian can be easily determined by the BIS emerged on the time-averaged spin polarization (TASP). Moreover,  one can  extend the dBSC to the high-order dBSC. Based upon the dimension reduction approach induced by high-order dBSC, Liu and co-workers have performed a series of studies in sudden quench regime \cite{2018Sci, experiment1, experiment2, experiment3, PRX2021, 2019charge, 2019general, 2021highcharge, 2021tenfold}, showing that the bulk topology of a $d$D phase can be uniquely characterized by the gradients of the TASP on arbitrary high-order BIS.  The highest order BIS are characterized by only discrete signs in the zero dimension, which greatly simplifies the characterization and detection of the equilibrium topological phases. As made evident on the platforms of optical lattice, the TASP can be obtained in  ultracold atomic system by spin population difference measured by spin-resolved time-of-light absorption imaging \cite{experiment1, experiment2, experiment3, Liu2013, Liu2014}. 

	These characterization schemes, however, are based on the ideal limit of sudden quench.  Under a more general dynamical setting with the system being slowly quenched with some finite quenching rate, one would expect more interesting physical behaviors and perhaps more efficient characterization schemes. In a previous work\cite{2020us}, it has been found that, nonadiabatic slow quench dynamics  can indeed provide a more efficient scheme in characterizing the topological phases. However, the previous study considered only a special class (${\mathbb{Z}}$-type)  of ten-fold classes of topological phases. It would be necessary and also interesting to extend the study to other types of topological phases. Compared with the sudden quench,  in addition to the BIS, a so-called spin inversion surface (SIS) in the momentum spaces was identified under slow quench dynamics. However, its underling physics and application remains to be explored. Moreover, the study of higher dimensional topological phases was only based on numerical results, and the exact solutions are still absent.

  To this end,  we first propose an exactly solvable multi-state Landau-Zener (LZ) model  that can be directly applied to the nonadiabatic slow quench dynamics of high-dimensional topological phases (including 3D ${\mathbb{Z}}$-type and  $\mathbb{Z}_2$-type topological phases). By analyzing the corresponding TASP of each model after quenching, we show that, the bulk topology of the system can be described not only by the values of TASP on order BIS, but also by the  gradients of TASP on order SIS. In addition, by introducing an extra quenching process, the topological invariant of the system can be directly described both by the values of TASP on order BIS and order SIS. Finally, from a dimensional reduction approach, we define the corresponding $n$-order SIS (BIS) in slow adiabatic quench dynamics, which quantifies a $d$D topological phase confined on the $(n-1)$-order SIS (BIS). The direct characterization of topological phases relying on arbitrary order BIS and SIS becomes possible. The above two schemes can  also be applied to characterize the  $\mathbb{Z}_2$-type topological phases. All the characterization schemes and their differences are summarized in  table~\ref{table:comparision}. Our findings complement the understanding of slow nonadiabatic dynamics and provide an alternative method to characterize the band topology for future experiments.

The paper is organized as follows. In Sec. II , we  present the explicit analytical solution of a four-state Landau-Zener problem that can be directly applied to the nonadiabatic dynamics of topological phases, and give an introduction to the dynamical field on both BIS and SIS in the slow nonadiabatic quench dynamics. Then in Sec. III, we present the results of the nonadiabatic dynamical characterization scheme  with single quenching process. Then In Sec. IV, we show  another scheme with more than one quenching process. In addition, the above two schemes are applied to characterize the  $\mathbb{Z}_2$-type topological phases in Sec. V. Finally, we provide a brief discussion and summary to the main results of the paper. 

\section{GENERIC THEORY IN quantum SLOW QUENCH dynamics}

In previous work \cite{2020us}, a $g/t$ version of two-level LZ model was exactly solved and applied to the study of  dynamical characterization of topological phases under slow quench. However, this two-level LZ model can only describe the dynamics of 1D and 2D $\mathbb{Z}$ topological systems, of which the minimal representation is written in terms of $2$ by $2$ matrices.   In this section, we  propose an exactly solvable multi-state LZ model that is closed related to higher-dimensional topological phases, whose matrix representation is $4$ by $4$ or even larger. Specifically, for a 3D  $\mathbb{Z}$-type insulator, or  $\mathbb{Z}_2$  topological insulator, its Hamiltonian is usually expressed in terms of Gamma matrices,  ${\cal H}(\bk) = \vec{h}(\bk)\cdot \vec{\gamma} = \sum_{j=0}^d h_{j}(\bk) \gamma_{j}$. Here, $\gamma_j$ are $4$ by $4$ matrices and obey the Clifford algebra $\{ \gamma_j, \gamma_l\} = 2 \delta_{jl}$ for $j, l = 0, 1, \ldots, d$. $d=3$ for 3D  $\mathbb{Z}$-type  insulator, and $d=4$ for  $\mathbb{Z}_2$ topological insulator. Without loss of generality, we can choose the following convention for Gamma matrices: $\gamma_0=\sigma_0\otimes \sigma_z$, $\gamma_{1, 2, 3}=\sigma_x\otimes \sigma_{x, y, z}$, $\gamma_4= \sigma_x \otimes \sigma_0$, with $\sigma_{x, y, z}$ the three Pauli matrices and $\sigma_0$ the $2$ by $2$ identity matrix.  To study the nonadiabatic dynamics of topological phases, we propose the following time-dependent multi-state Landau-Zener problem:
\be
{\cal H} (t) = \Big( h_0 + \frac{g}{t} \Big) \gamma_0 + \sum_{j=1}^4 h_j \gamma_j.
\ee
Or, in matrix form: 
\be
{\cal H} (t)  = \left( \begin{array}{cccc}
h_0 + g/t & 0 & h_3-ih_4 & h_1- i h_2  \\
0 & h_0+ g/t & h_1+ ih_2 & -h_3-i h_4  \\
 h_3+ih_4 & h_1- i h_2 &-h_0- g/t  & 0  \\
 h_1 +ih_2  & -h_3+i h_4 & 0 & -h_0 -g/t
\end{array}
\right) .\nn \\ \label{eq:LZ}
\ee
The parameter $g$ determines the  quenching rate. It varies from $0$ to $\infty$, corresponding to a continuous crossover from the sudden quench limit ($g=0$) to adiabatic limit ($g\rar \infty$). The form of $g/t$ enables us to quench the system from initially trivial Hamiltonian at $t=0^+$ to a final topological state at $t=\infty$ by keeping the other parameters unchanged, and at the same time obtain an analytical transition probability. 

One needs to find the evolution of state vector $|\Psi(t)\ra$ governed by the time-dependent Schrodinger equation: $i\hbar \frac{d}{dt}|\Psi(t)\ra = {\cal H}(t) | \Psi(t)\ra$. 
The traditional Landau-Zener problem is to find the transition probabilities from an initial state to some final energy state. The exactly solvable models are rare and have been found to exist only in some special forms \cite{sinitsyn-14pra, sinitsyn16pra, li17pra, li18prl}. Note that a four-state LZ problem with path interference was also studied \cite{Sinitsyn2015}, but is distinctively different from Eq.~(\ref{eq:LZ}). Moreover, in our problem, we  need to find the amplitude of state vector after nonadiabatic transition. 

Under Hamiltonian (\ref{eq:LZ}), our purpose is to look for the final state vector at time $t\rar \infty$ starting from the initial state $|\Psi(0^+)\ra$ at $t\rar 0^+$.  At initial time $t\rar 0^+$, the time-dependent term $g/t$ is infinitely large compared with other parameters, and thus the Hamiltonian is trivial. The initial instantaneous eigenenergies are given by $E=\pm g/t$ which are infinitely large, with eigenstates $(1,0, 0, 0)^T$, $(0,1, 0, 0)^T$, $(0,0, 1, 0)^T$ and $(0,0, 0, 1)^T$, respectively. At final time $t\rar \infty$, the time-dependent term $g/t$ goes to zero, and the final Hamiltonian ${\cal H}(t\rar \infty)$ has  instantaneous eigenenergis $E=\pm \ve = \pm \sqrt{\sum_{j=0}^4 h_j^2}$, both of which are doubly degenerate.  The eigenstates are written as: 
\be
|\psi_{+1}\ra &=& {\cal A}^{-1} (\ve + h_0, 0,h'_+ , h_+)^T,\\
|\psi_{+2}\ra &=&{\cal A}^{-1} (0,\ve + h_0, h_- , -h'_- )^T,
\ee
for eigenvalues  $E=+\ve$, and 
\be
|\psi_{-1}\ra &=&{\cal A}^{-1} (- h'_-, -h_+,\ve + h_0, 0)^T,\\
|\psi_{-2}\ra &=&{\cal A} ^{-1}(-h_-, h'_+, 0 ,\ve + h_0 )^T,
\ee
for eigenvalues of $E=-\ve$.   Here we introduce $h_{\pm} = h_1 \pm h_2$,  $h'_{\pm} = h_3 \pm h_4$, and normalization factor ${\cal A} ={\sqrt{2\ve (\ve + h_0)}}$.

If we prepare the system in its initial ground state,  the system will undergo a nonadiabatic transition during the evolution, and finally at time $t\rar \infty$, the system will stay not only on the final ground state, but also on the excited state  with nonzero probability.  It can be shown that (details are given in Appendix A), starting from the initial state $|\Psi(t\rar 0^+) = (0, 0, 1, 0)^T$, the final wave function at $t\rar \infty$ is 
\be
|\Psi(t) \ra =&& \sqrt{P_u}e^{-i\ve t -i\delta} \left(\frac{h'_+}{h} | \psi_{+1} \ra+ \frac{h_+}{h} | \psi_{+2} \ra \right)  \nn \\
&&  + \sqrt{P_d} e^{i\ve t} |  \psi_{-1} \ra,  \label{eq:psit}
\ee
with the transition probability 
\be
P_d = 1-P_u=\frac{e^{2\pi g} - e^{-2\pi g \frac{h_0}{\ve}}}{e^{2\pi g} - e^{-2\pi g}}.
\ee
  Here, we introduce $h= \sqrt{\sum_{j=1}^4 h_j^2} $, and $\delta$ as an undetermined phase factor. 

Note that the Hamiltonian (\ref{eq:LZ}) possesses a symmetry ${\cal P} {\cal H} {\cal P}^{-1}= {\cal H}$ with ${\cal P} =  i\sigma_z \otimes \sigma_y \hat{K}$. Here, ${\cal P}$ is an anti-unitary operator with $\hat{K}$ being the complex-conjugate operator that forms the complex conjugate of any coefficient that multiplies a ket.  This discrete symmetry is the reason why the eigenenergies are always doubly degenerate, which is similar to the Kramers degeneracy. Operator ${\cal P}$ transforms one eigenstate to its degenerate partner up to a phase factor, e.g., $|\psi_{-2} \ra = {\cal P} |\psi_{-1} \ra $. Therefore, the final state starting from initial state $ (0, 0, 0, 1)^T$ is simply given by performing ${\cal P}$ on $ |\Psi(t) \ra$ of Eq.~(\ref{eq:psit}) .

The dynamics of spin polarization $\la {\gamma}_i (t) \ra \equiv \la \Psi(t) | {\gamma}_i| \Psi(t)\ra$, and the final TASP over a period $T$ can be further obtained:
\be
\overline{\la {\gamma}_i\ra} = (P_u-P_d) \frac{h_i}{\ve}.
\label{eq:spin polarization}
\ee 
This result tells us the vanishing spin polarization in each component of TASP can have two different cases: one is $h_i=0$, the other is $(P_u-P_d)=0$. The latter is defined as SIS, and its position is dependent on parameter $g$.  On the SIS, all the components of TASP $\overline{\la \gamma_{i} \ra}$ vanish. One needs to define a new field  ${\widetilde{{g_{so,i}}}}=-_{\partial{k_{\perp}}}{\overline{\la \gamma_{so,i}\ra}}$ to characterize topology, which can be shown to be proportional to the normalized spin-orbit field $\hat{h}_{so,i}$. Here, spin-orbit field means the remaining components of $\bm h$ except $h_0$. 

The position of BIS is given by $h_0=0$, and is independent of driving factor $g$.  As one increases $g$, the separation between BIS and SIS  is enlarged, and they totally overlap when $g$ equals to zero. On the BIS,  we see from (\ref{eq:spin polarization}) that 
 $\overline{\la { {\gamma}_{so,i}}\ra} = c \hat{h}_{so,i}$ with prefactor $c= (1-\cosh2\pi g)/\sinh 2\pi g$ which is usually negative. Thus, one can also characterize the topology by simply  measuring the value of  TASP.
  
\begin{figure}[htbp]
	\centering
	\epsfig{file=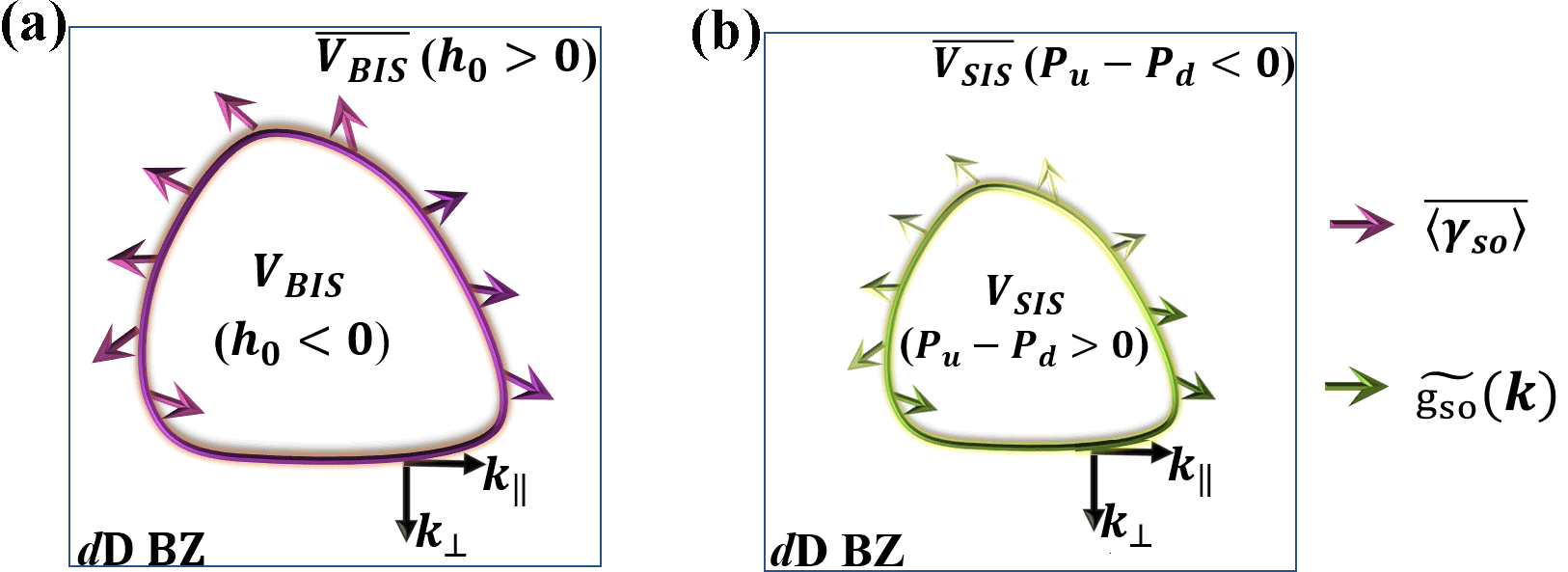, width=3.0in}
	\caption{Schematic diagram of the topological characterization in the slow nonadiabatic quench dynamics after only quenching $h_0$ axis. The $d$D topological phases can be characterized by the winding of the corresponding dynamical field (colored arrows) on both BIS and SIS. Here the momentum space is decomposed into $k_\perp$ and  $k_\Vert$ (black arrows) with $k_\perp$ perpendicular to the $V_{\tiny{BIS,SIS}}$ and pointing to the complementary region of $V_{\tiny{BIS,SIS}}$.}
	\label{fig:BIS_SIS}
\end{figure}

  In short, based on the fact that the topological invariant can be obtained by winding of $\bm{{\hat h}_{so}}$ both on BIS and SIS, one can easily capture the bulk topology from the value or the gradient of TASP in the experiment. For clarity, we give a schematic diagram of the topological characterization in the slow nonadiabatic quench dynamics, as shown in Fig.~\ref{fig:BIS_SIS}. The information related to the BIS and SIS is always marked in purple and green in the following description, respectively.

\section{the first dynamical characterization scheme}
After extending the analytical solution of TASP to the high-dimensional topological phases, we turn to exploring the dynamical characterization scheme with different topological models. The first dynamical characterization scheme includes two steps: (i) Performing a deep quench along a certain direction to obtain all the spin-textures; (ii) Characterizing the bulk topology in the same quenching process. The detailed processes are shown below.

{\it 1D case:} To exemplify, we illustrate our scheme by considering the 1D topological phases of AIII class with Hamiltonian spanned by two Pauli matrices ${\cal H}(k_x) = \bm h(k_x) \cdot {\bm \sigma}$. We choose the quenching axis to be along $z$ direction. 
\be
&&h_{so} \equiv h_x(k_x) = t_{so} \sin k_x ,\nonumber \\ 
&&h_0 \equiv  h_z(k_x) =  m_z- t_0\cos k_x .
\ee
When $h_z$ is quenched from infinity at $t=0^+$, to the final value  at $t=\infty$, the system is tuned from a trivial phase to a phase that depends on the value of $m_z$. If $|m_z| > t_0$, the post-quench Hamiltonian is still in the trivial phases, while if $|m_z| < t_0$, the post-quench Hamiltonian is in the  nontrivial phase.  The equilibrium version of this model has been realized in ultracold atomic system \cite{Song2018}. 

\begin{figure}[htbp]
	\centering
	\epsfig{file=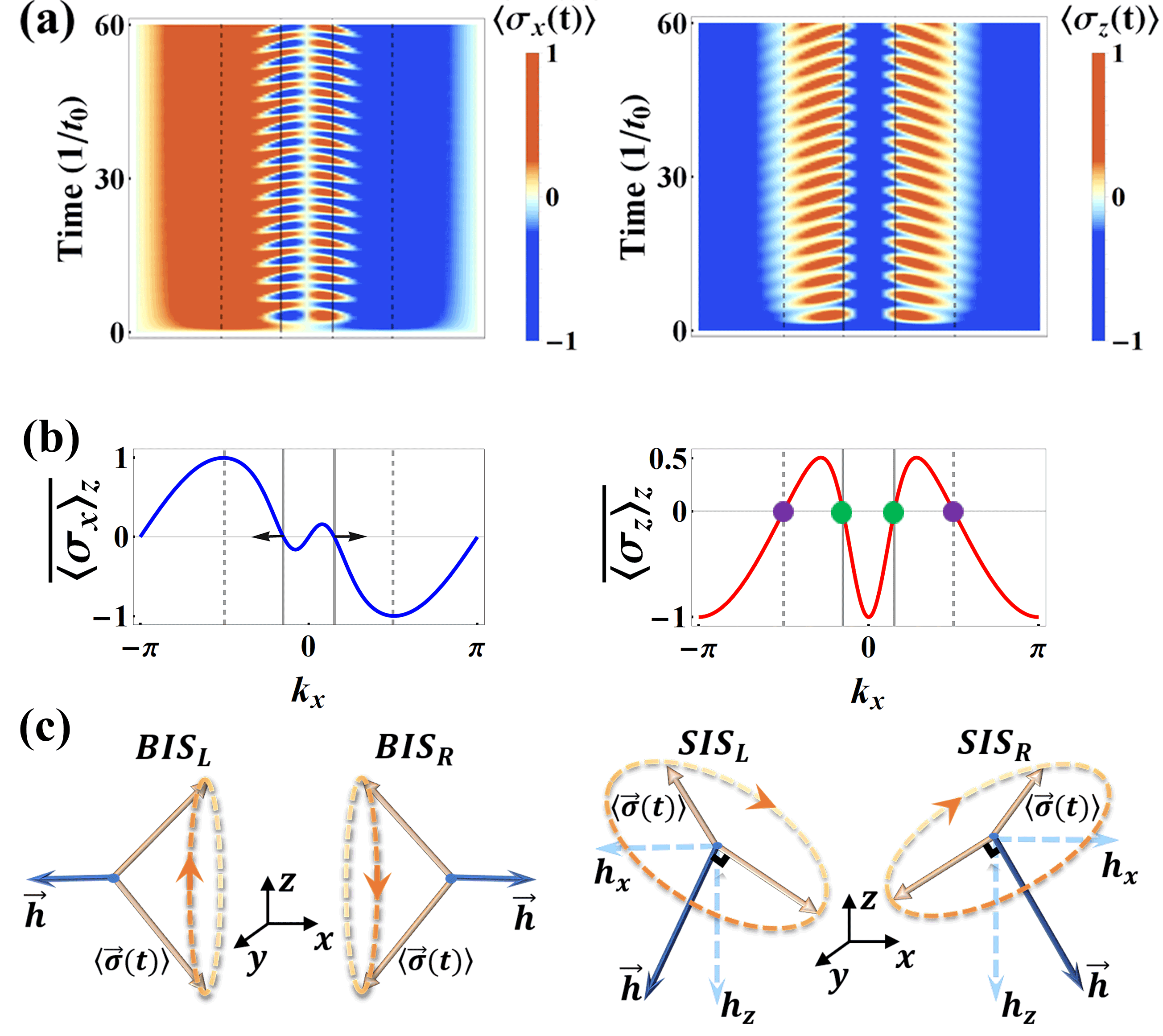, width=3.0in}
	\caption{ (a) Time evolution and (b) the time-averaged spin polarization  of 1D topological model are shown after a slow quench with $t_{so} =t_0=g=1$, $m_z=0$. $\overline{\la{ \sigma}_{x}\ra}_z$ represent the $x$-component of TASP after quenching $h_z$ axis. The dashed lines denote the BIS points. We define the points of solid lines as spin inversion surface (SIS), on which all components of spin polarization are vanishing. They are also indicated by green points and purple points in $\overline{\la{ \sigma}_{z}\ra}_z$.  The corresponding dynamical field can be determined by the opposite signs of $x$-component $\overline{\la { \sigma}_{x}\ra}$ on the BIS and the negative gradient ${\widetilde{{g_{x}}}}=-_{\partial{k_{\perp}}}{\overline{\la \sigma_{x}\ra}}$ on the SIS, and thus gives the nontrivial winding number $1$. (c) the approximately stable dynamic precession  behavior of spin polarization on BIS and SIS after a certain time.}
	\label{fig:1D_Z}
\end{figure}

In Fig.~\ref{fig:1D_Z} (a) and (b), we present the results of time evolution of spin polarization and TASP of 1D topological model after {\it only slowly quenching $h_z$ axis}. The spin at each $k_x$ point processes after quenching system from a fully spin-polarized state pointing along negative $h_0$ axis. We can observe an approximately stable dynamic precession  behavior of spin polarization  on both BIS (dashed lines) and SIS (solid lines). For BIS point  with $k_x=\frac{\pi}{2}$,  the field along $x$ direction ($h_x \equiv 1$) becomes much larger than the the field along $z$ direction ($g/t$) after a certain time $\Delta t$.  ${\la \sigma_x(t) \ra}$ is always a nonzero value while ${\la \sigma_z(t) \ra}$ oscillates between positive  and negative value. Fig.~\ref{fig:1D_Z} (c) vividly illustrates the above dynamical behavior of spin polarization. On the two BISs (purple points),  ${\la \bm{\sigma}(t) \ra}$ precesses around the field $h_x$, leading to vanishing values for two components  of TASP ${\overline{\la \sigma_y \ra}}$ and ${\overline{\la \sigma_z \ra}}$. Another component of TASP ${\overline{\la \sigma_x \ra}}$  is nonzero at the two BIS points, but shows opposite signs. 
 
 Moreover, in addition to the characterization on BIS in our previous paper \cite{2020us}, we highlight another type of zeros appearing in $\overline{\la { \sigma}_{z}\ra}$ on the SIS as indicated by green points in Fig.~\ref{fig:1D_Z}(b). On the SIS, both ${\la \sigma_x(t) \ra}$ and ${\la \sigma_z(t) \ra}$ undergo a resonant spin-reversing transitions. Therefore, three components of TASP vanish: $\overline{\la { \sigma}_{x}\ra}=\overline{\la { \sigma}_{y}\ra}=\overline{\la { \sigma}_{z}\ra}=0$, from which, one can conclude that the effective field  is perpendicular to the spin polarization  ${\la \bm{\sigma}(t) \ra}$ as shown in Fig.~\ref{fig:1D_Z}(c). 
 One may cannot use the value of TASP on SIS to characterize the topology. Surprisingly, in addition to BIS, we can also use the dynamical gradient field ${\widetilde{{g_{x}}}}=-_{\partial{k_{\perp}}}{\overline{\la \sigma_{x}\ra}}$ on the SIS, which gives rise to a nontrivial winding number ${\cal{W}}$=$\left|{\frac{1}{2}[\sgn({\widetilde{{g_{x}}}}_{k_{x,-}})-\sgn({\widetilde{{g_{x}}}}_{k_{x,+}})]}\right|=1$, to characterize the topology, as shown by the antiparallel black arrows in Fig.~\ref{fig:1D_Z} (b). Compared to the sudden quench approach in which one has to characterize the topology  relying only on BIS, our slow nonadiabatic quench approach provides two complementary schemes for characterizing topology through the TASP defined not only on BIS but also on SIS. 

 Note that in 1D case, the topological invariant is captured by 0D points on 1-BIS and 1-SIS with different signs, which remind us a $d$D bulk topological phase may be also characterized by 0D points defined on the high-order BIS or SIS. We will show this point in the case described below. By the way, the most common 1D Su-Schrieffer-Heeger model, which belongs to the BDI classes, can also be characterized by the same method.

\begin{figure}[htbp]
	\centering
	\epsfig{file=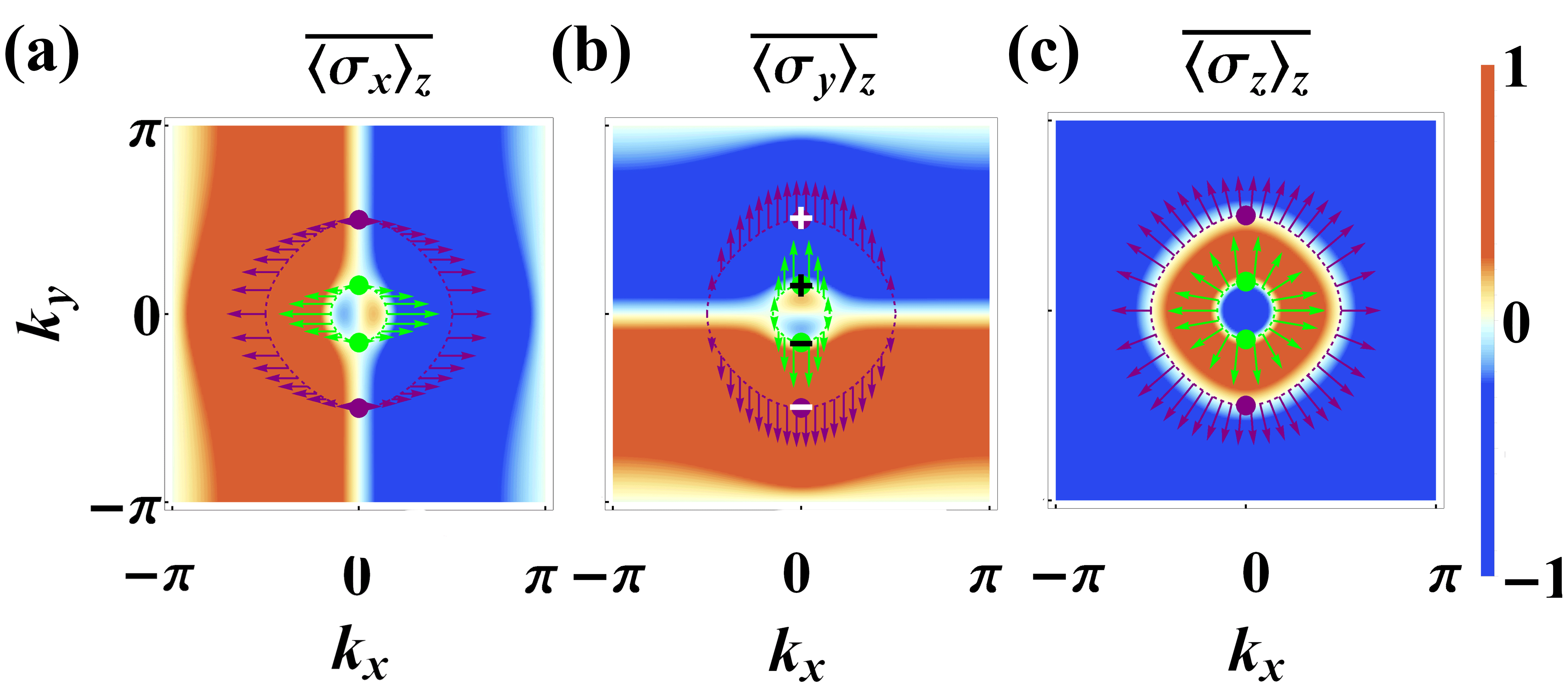, width=3.4in}
	\caption{Analytical results for 2D Chern insulators with $m_z = t_0 =t_{so}^{x,y}=g=1$,$m_{x}=m_{y}=0$. (a)(b)(c) TASP are plotted after slowly quenching $h_z$ axis. 1-SIS is the surface with all the three components vanishing $\overline{\la {\bm \sigma} \ra}=0$ (green dashed ring) while on 1-BIS, $\overline{\la {\sigma}_{z} \ra}=0$ but $\overline{\la {\sigma}_{x,y} \ra}\neq0$ (purple dashed ring). The purple and green arrows are the vectors formed by $-\overline{\la \sigma_{x,y} \ra}$ and ${\widetilde{{g_{x,y}}}}$, both indicating a nontrivial topological pattern with  Chern number ${\cal C}_{1}=-1$. In the meanwhile, 2-BIS and 2-SIS denoted by purple and green points, respectively, also give the same Chern number.}
	\label{fig:2D_Z}
\end{figure}

{\it 2D case:} Now we apply the first characterization scheme to study 2D topological phases, which are generically described by a two-band Hamiltonian spanned by three Pauli matrices: ${\cal H}(\bk) = \bm h(\bk) \cdot {\bm \sigma}$ , with the vector field given by:
\be
&&h_x=m_x+t_{so}^x \sin k_x, \label{eq:2Dhx}\nonumber\\
&&h_y = m_y + t_{so}^y  \sin k_y, \label{eq:2Dhy} \\
&&h_z = m_z - t_0 \cos k_x - t_0 \cos k_y\nonumber \label{eq:2Dhz}
\ee
This Hamiltonian has been realized in recent experiment of quantum anomalous Hall effect \cite{Wu2016}. Here we slowly quench the $z$-component of vector field from $t=0^+$ to $\infty$, and study the topological characterization of post-quench Hamiltonian determined by $m_z$. For $0<m_{z}<2t_0$, the post-quench Hamiltonian gives a topological phase with Chern number ${\cal C}_{1}=-1$. For $-2t_0 < m_{z}<0$, the post-quench system describes a topologically nontrivial phase with Chern number ${\cal C}_{1}=+1$.

In Fig.~(\ref{fig:2D_Z}) (a), (b), and (c), we plot the three components of the TASP {\it by only quenching the $h_z$ axis}.  The vanishing spin polarization  in  $\overline{\la { \sigma}_{z} \ra}$ are denoted by the green dashed ring and purple dashed ring, respectively. On the green dashed ring, the other two components  $\overline{\la { \sigma}_{x} \ra}$  and $\overline{\la { \sigma}_{y} \ra}$ also vanish, and thus we identify it as first-order SIS (1-SIS).  The purple dashed ring is identified as the first-order BIS (1-BIS),  on which the other two components $\overline{\la { \sigma}_{x} \ra}$  and $\overline{\la { \sigma}_{y} \ra}$ are nonzero.  Similar to the 1D case, one can determine the topological invariant of post-quench Hamiltonian from the topological pattern formed by the other two components $-\overline{\la { \sigma}_{x,y} \ra}$ on 1-BIS, or by the gradients of  two other components ${\widetilde{{g_{x,y}}}}$ on 1-SIS. As shown in Fig.~(\ref{fig:2D_Z}) (a) and (b), we plot the non-zero values of spin polarization $-\overline{\la { \sigma}_{x} \ra}$  and $-\overline{\la { \sigma}_{y} \ra}$ on the 1-BIS by the purple arrows, and combine them in the plot of $\overline{\la { \sigma}_{z} \ra}$. From the combined purple arrows, i.e, the dynamical field on TASP, one can easily determine that the post-quench Hamiltonian is topologically non-trivial with topological invariant ${\cal C}_{1}=-1$. Moreover, we also plot the gradients ${\widetilde{{g_{x,y}}}}$ on the 1-SIS by green arrows, and combine them in the density plot of $\overline{\la { \sigma}_{z} \ra}$. One can see that the green arrows show a same topological pattern as the purple ones, and thus serves as an alternative scheme in characterizing the topological invariant of post-quench Hamiltonian.  

Now we further study the high-order characterization of BIS and SIS. We observe that, in Fig.~(\ref{fig:2D_Z}) (a), (b), there exist two white lines on which the corresponding components of spin polarization are zero. We define the  second-order BIS (2-BIS)  and  second-order SIS (2-SIS) as the intersection points of the white lines in $\overline{\la { \sigma}_{x}\ra}$ and the 1-BIS and 1-SIS, respectively, i.e., $\overline{\la \sigma_{x} \ra}=0\cap\overline{\la \sigma_{z} \ra}=0$, as indicated by the purple and green points shown in Fig~(\ref{fig:2D_Z}) (a-c). 
One can then determine the topology of post-quench Hamiltonian by the topological pattern on 2-BIS or 2-SIS. Specifically, on the 2-BIS points, one finds that the values of $-\overline{\la{ \sigma}_{y}\ra}$ are $+1$ at positive $k_y$ point and $-1$ at negative $k_y$ point, as marked by the white opposite signs $+$ and $-$ in $\overline{\la{ \sigma}_{y}\ra}$. The nontrivial topological invariant of the original 2D post-quench Hamiltonian can then be obtained by ${\cal C}_{1}$=${\frac{1}{2}[\sgn({-\overline{\la{ \sigma}_{y}\ra}}_{k_{y,-}})-\sgn({-\overline{\la{ \sigma}_{y}\ra}}_{k_{y,+}})]}=-1$. On the 2-SIS points, one can also find that the gradients ${\widetilde{{g_{y}}}}$  are $+1$ at positive $k_y$ point and $-1$ at negative $k_y$ point, as marked by the black opposite signs $+$ and $-$ in $\overline{\la{ \sigma}_{y}\ra}$.  The value of topological invariant can then be obtained by ${\cal C}_{1}$=${\frac{1}{2}[\sgn({\widetilde{{g_{y}}}}_{k_{y,-}})-\sgn({\widetilde{{g_{y}}}}_{k_{y,+}})]}=-1$. The topological characterization of these 0D high-order points coincides with 0D points in 1D case. By the way, as discussed in appendix B, the method of constructing (high-order) BIS and SIS from the TASP is also consistent with the definition of (high-order) BIS and SIS. 

\begin{figure}[htbp]
	\centering
	\epsfig{file=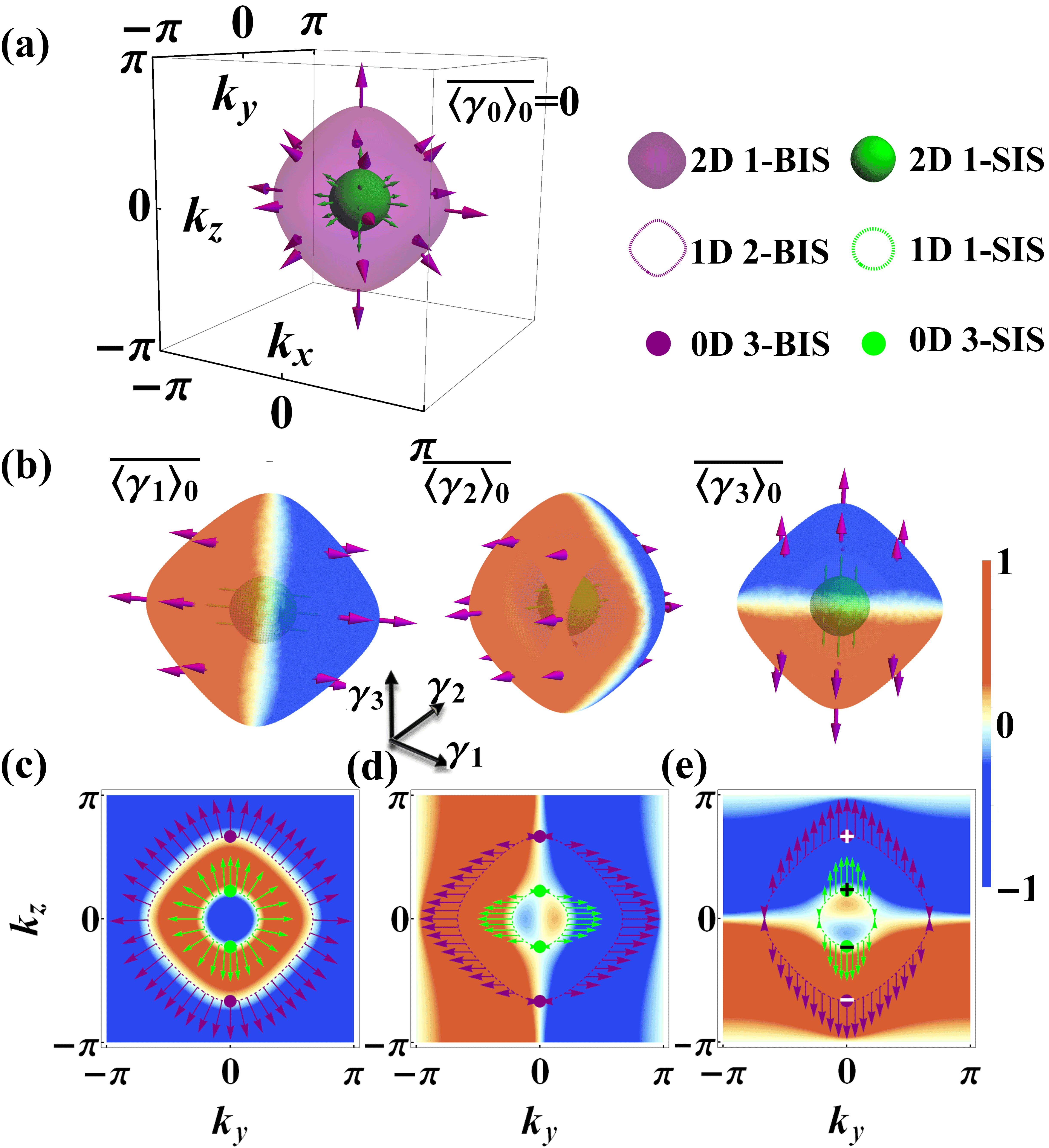, width=3.4in}
	\caption{Analytical results for topological characterization of 3D chiral topological phases by quenching $h_0$ axis. Here we set $t_{so}=t_{0}=g=1,m_z=1.5$. (a) The 1-BIS (purple surface) defined by $\overline{\la \gamma_{0} \ra}=0$ and the topological dynamical field $-\overline{\la {\bm \gamma_{so} } \ra}$ (purple arrows) composed by the three components $-\overline{\la \gamma_{1,2,3} \ra}$ (purple arrows on the three surfaces in (b)). For 1-SIS contained within 1-BIS, the situation is similar, except that the topological dynamical becomes the gradient  instead of the spin polarization. (b) The three components of TASP $\overline{\la \gamma_{1,2,3} \ra}$ on 1-BIS and 1-SIS. Their values of  $-\overline{\la \gamma_{1,2,3} \ra}$ and ${\widetilde{{g_{1,2,3}}}}$ are illustrated by purple and green arrows, respectively. (c)  the 2-BIS and 2-SIS (purple dashed ring and green dashed ring) constructed by the intersection of $\overline{\la \gamma_{0} \ra}$ and $\overline{\la \gamma_{1} \ra}$  on cross sections $k_{x}=0$. (d)(e): the two components of TASP $\overline{\la \gamma_{2,3} \ra}$ on cross sections $k_{x}=0$.  Topological dynamical field $-\overline{\la \gamma_{2,3} \ra}$ and ${\widetilde{{g_{2,3}}}}$ are still denoted by purple and green arrows, respectively. In the meanwhile, 3-BIS and 3-SIS denoted by purple and green points, respectively, also correspond to the same winding number. }
	\label{fig:3D_Z}
\end{figure}

{\it 3D case:} In order to show the first characterization scheme is not limited to 1D and 2D cases, we further study the topological characterization of high dimensional topological phases under slow nonadiabatic quench.
The Hamiltonian of the 3D topological phases can be written as: $H(\bk) = \sum_{j=0}^3 h_j(\bk) \gamma_j$, with
\be
&&h_0=m_z -t_0 \sum_{i= x,y,z} \cos k_i,\nonumber\\ 
&& h_{1,2,3}= t_{so} \sin k_{x,y,z}.
\ee
 We start the slow quench from time $t=0^+$ so that the system is initially in a trivial phase. At time $t\rar \infty$, the system may lie in different phases depending on the values of $m_z$: for $|m_z|>3t_0$, a trivial phase; for $t_0<m_z <3t_0$, a topological phase with winding number $\nu_3=-1$; for  $-t_0 < m_z<t_0$, with $\nu_3=2$; and for $-3t_0< m_z< -t_0$, with $\nu_3=-1$. 

 As plotted in Fig.~\ref{fig:3D_Z} (a), one can see that, after  {\it only slowly quenching the $h_0$ axis}, both 1-BIS and 1-SIS (contained within 1-BIS) appear in $\overline{\la \gamma_{0} \ra}$ as closed surfaces where spin polarization vanished, as indicated by the purple and green surfaces. As described in the 1D and 2D cases, the topological invariant of the post-quench Hamiltonian can be determined by other three spin texture components  $-\overline{\la \gamma_{1,2,3} \ra}$ (purple arrows) on 1-BIS and ${\widetilde{{g_{1,2,3}}}}$ (green arrows) on 1-SIS, respectively. As shown in Fig.~\ref{fig:3D_Z} (b),  the non-zero values of spin polarization $-\overline{\la { \gamma}_{1} \ra}$,  $-\overline{\la { \gamma}_{2} \ra}$, and $-\overline{\la { \gamma}_{3} \ra}$ are plotted by the purple arrows, and combined  on the 1-BIS in $\overline{\la { \gamma}_{0} \ra}$. The above combined purple arrows show a 3D topological pattern cover over the surface of 1-BIS, and thus give a nontrivial winding number $\nu_{3}=-1$. Moreover, topological characterization on 1-SIS is the same as that on 1-BIS except that spin polarization $-\overline{\la \gamma_{1,2,3} \ra}$ is replaced by gradient ${\widetilde{{g_{1,2,3}}}}$.
 
 For second-order BIS and SIS, we should further consider another component of spin polarization, taking $\overline{\la \gamma_{1} \ra}$ as an example, whose zeros are located on the plane with $k_x=0$ as given by $h_1=0$. By finding the intersection : $\overline{\la \gamma_{0} \ra}=0\cap\overline{\la \gamma_{1} \ra}=0$ , we then obtain the 2-BIS (purple dashed ring) and 2-SIS (green dashed ring) on the 1-BIS and 1-SIS, as presented in Fig.~\ref{fig:3D_Z}(c).  In addition, the other two components  $\overline{\la \gamma_{2,3} \ra}$ on the plane $k_{x}=0$ are also obtained,  and  the values of  $-\overline{\la \gamma_{2,3} \ra}$ on 2-BIS and ${\widetilde{{g_{2,3}}}}$ on 2-SIS are also illustrated by the purple and green arrows, respectively,  in Fig.~\ref{fig:3D_Z} (d) and (e). Compared  with the 2D case, the combined arrows on the 2-BIS and 2-SIS then give the same  Chern number ${\cal C}_{1}=-1$,  corresponding to nontrivial topological 3D phases with winding number $\nu_3=-1$.  
 Furthermore, the 3-BIS and 3-SIS are constructed on 2-BIS and 2-SIS by the intersection: $\overline{\la \gamma_{0} \ra}=0\cap\overline{\la \gamma_{1} \ra}=0\cap\overline{\la \gamma_{2} \ra}=0$, as indicated by the purple and green points, respectively. Naturally, the opposite signs on the 3-BIS and 3-SIS then give the same nontrivial topological number ${\cal W}=-1$, which also corresponding to 3D phases with winding number $\nu_3=-1$.
 
Overall, in order to determine the information of $n$-BIS or $n$-SIS, we only need to consider the spin polarization on the $(n-1)$-BIS or SIS, rather than the whole BZ. More importantly, the highest-order BIS or SIS of system are only pairs of 0D points with different signs. Thus, the characterization with high-order BIS or SIS will greatly simplifies the strategy of topological characterization compared with low-order BIS or SIS. 

\section{the Second dynamical characterization scheme}
We proceed to introduce the second scheme to the characterization of the bulk topology of system. Compared with the first scheme, the second one need two different quenching processes to determine the bulk topological invariant. Specifically, (i) Performing a deep quench along a certain direction to determine the position of SIS, (ii) Further performing  a deep quench along another direction to capture the topology of the system on the above obtained SIS. To avoid redundancy, we no longer show the results of 1D case.

\begin{figure}[htbp]
	\centering
	\epsfig{file=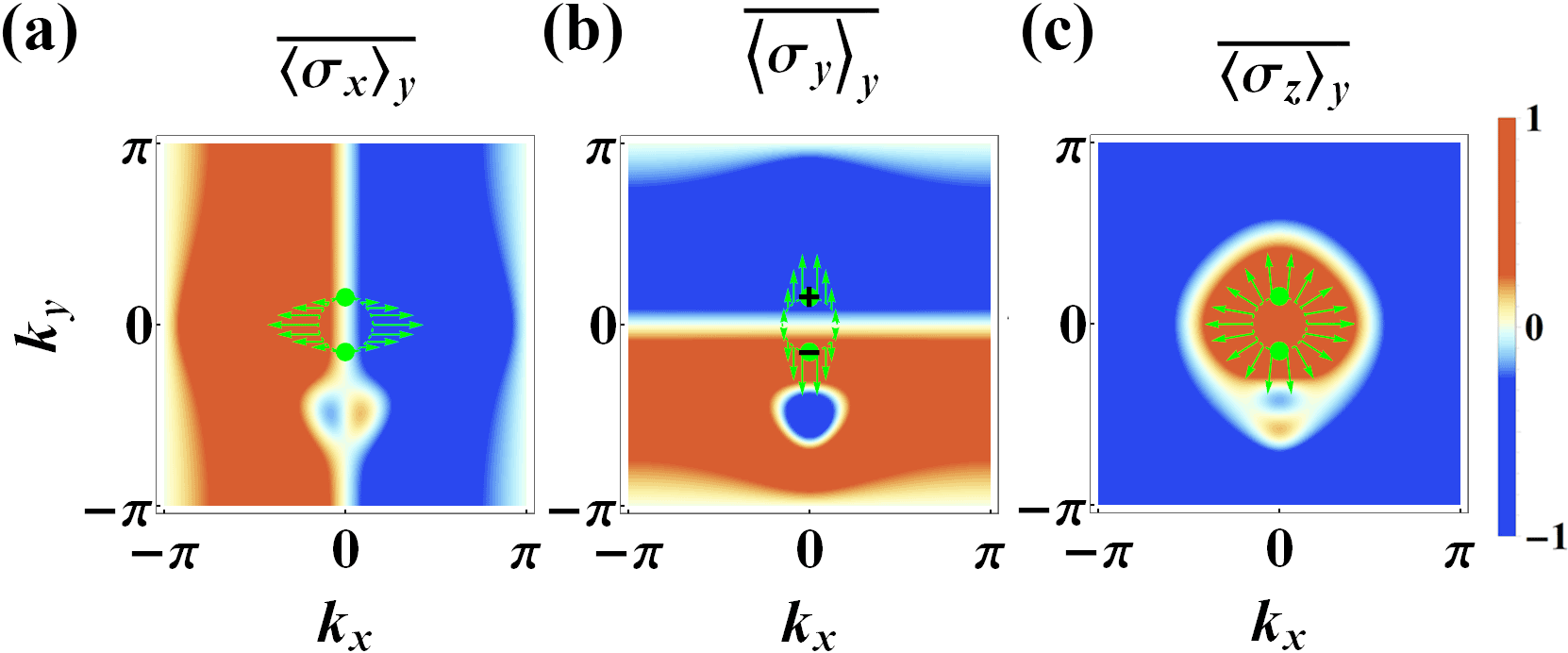, width=3.2in}
	\caption{The direct characterization based on SIS for 2D Chern insulator. The green line represents the SIS of slowly quenching  $h_z$ axis. The value of two spin textures $-\overline{\la {\sigma}_{x,y} \ra}_y$ on SIS can form a topological pattern corresponding to Chern number ${\cal C}_{1}=-1$. In the meanwhile,  2-SIS constructed by the 1-SIS and $\overline{\la {\sigma}_{x} \ra}_y=0$, also give the same Chern number. }
	\label{fig:SIS2D_second}
\end{figure}

{\it 2D case:} We first deeply quench the $h_z$ axis with time from $0^+$ to $+\infty$. Then the 1-SIS with $\overline{\la {\bm \sigma} \ra}_z=0$ is obtained, which is indicated by the green ring in Fig.~\ref{fig:SIS2D_second}. We further perform a deep quench along another axis $h_y$, and measure the corresponding value of TASP on the above 1-SIS. A topological pattern is formed on 1-SIS by the value $-\overline{\la {\sigma}_{x,y} \ra}_y$ corresponding to Chern number ${\cal C}_{1}=-1$. In the meanwhile,  2-SIS constructed by the 1-SIS and $\overline{\la {\sigma}_{x} \ra}_y=0$, also gives the same Chern number, as the opposite signs shown in Fig.~\ref{fig:SIS2D_second} (b) . 

\begin{figure}[htbp]
	\centering
	\epsfig{file=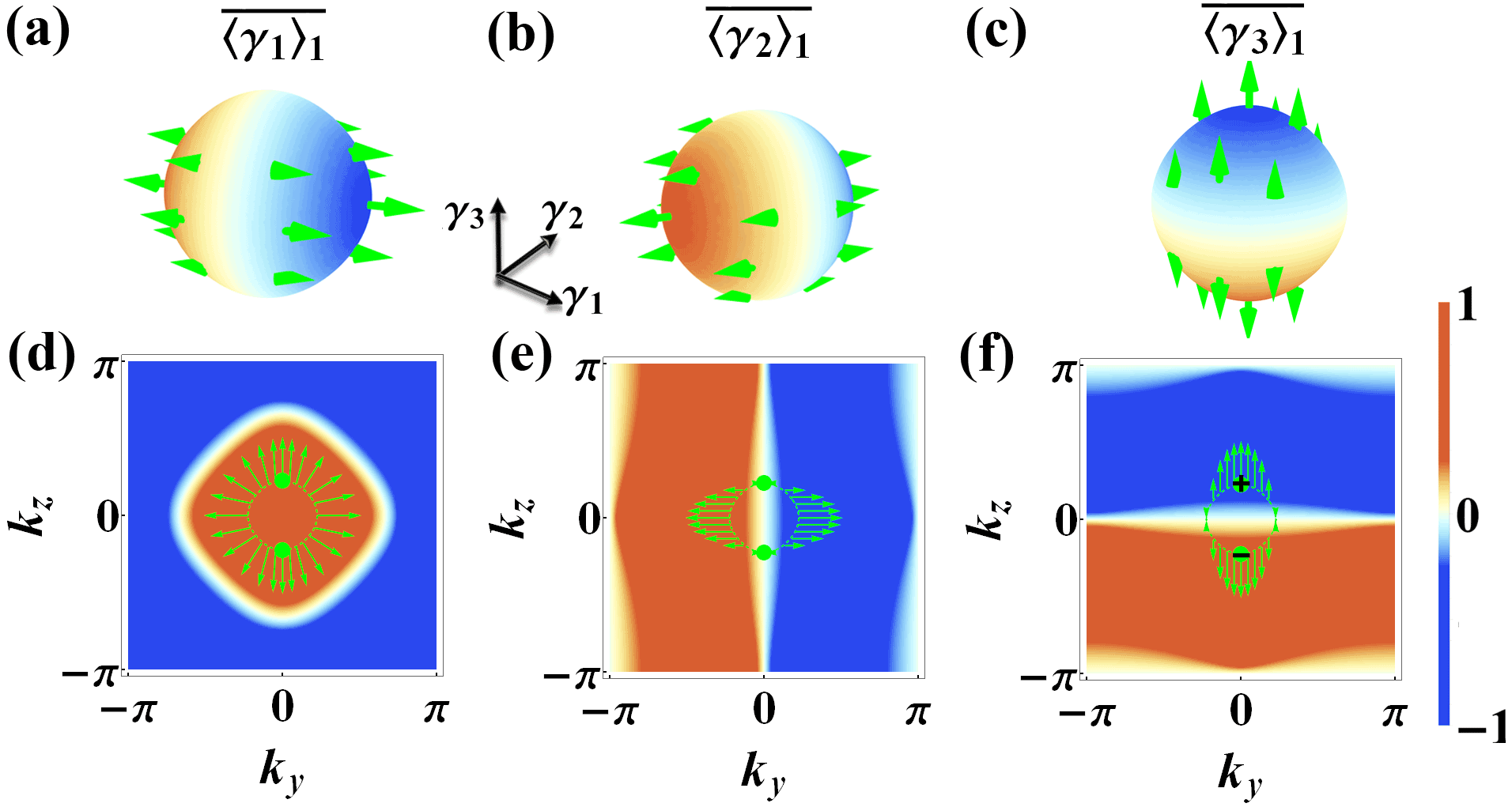, width=3.2in}
	\caption{The direct characterization based on SIS for 3D chiral insulator.(a-c) The corresponding value of TASP on the  1-SIS of slowly quenching $h_0$ axis. (d-f) The green line represents the 2-SIS of slowly quenching $h_0$ axis. The value of two spin textures in (e) and (f) on 2-SIS can form a topological pattern corresponding to Chern number ${\cal C}_{1}=-1$. In the meanwhile, 3-SIS constructed by the 2-SIS and $\overline{\la {\gamma}_{2} \ra}_1=0$, also gives the same topological number. }
	\label{fig:SIS3D_second}
\end{figure}

{\it 3D case:} We first deeply quench the $h_0$ axis, and then the 1-SIS with $\overline{\la {\bm \gamma} \ra}_0=0$ is obtained. 
 We further perform a deep quench along another axis $h_1$, and measure the corresponding value of TASP on the above 1-SIS, as shown in Fig~\ref{fig:SIS3D_second}(a-c). A topological pattern can be formed by the value of $-\overline{\la {\gamma}_{1,2,3} \ra}_1$ corresponding to winding number ${\cal \nu}_{3}=-1$.   For second-order SIS, we still take another component of spin polarization $\overline{\la \gamma_{1} \ra}_1$ as an example, whose zeros are located on the plane with $k_x=0$ as given by $h_1=0$. By finding the intersection : $\overline{\la \gamma_{0} \ra}_1=0\cap\overline{\la \gamma_{1} \ra}_1=0$ ,  the  2-SIS (green dashed ring)  is then obtained, as presented in Fig.~\ref{fig:SIS3D_second} (d-f). The values of  $-\overline{\la \gamma_{2,3} \ra}_1$ on 2-SIS are also illustrated by the green arrows,
  and their combined arrows on the 2-SIS then give the same  topological number $-1$ in Fig.~\ref{fig:SIS3D_second} (d). Moreover, the 3-SIS is constructed on 2-SIS by the intersection: $\overline{\la \gamma_{0} \ra}_1=0\cap\overline{\la \gamma_{1} \ra}_1=0\cap\overline{\la \gamma_{2} \ra}_1=0$, as indicated by green points. Similarly, the opposite signs on 3-SIS give the same nontrivial topological number ${\cal W}=-1$, which also corresponding to the winding number $\nu_3=-1$.

\section{the dynamical characterization of $\mathbb{Z}_2$-type topological phases }

{\it The first scheme:} Finally, we apply the above two schemes to the $\mathbb{Z}_2$-type topological phases. As an example, we choose 3D topological phases in class AII, which is characterized by Chern-Simons invariant $CS=\sgn[(m-3t_0){(m-t_0)}^3{(m+t_0)}^3(m+3t_0)]$ \cite {ten3d} with Hamiltonian $H(\bk) = \sum_{j=0}^4 h_j(\bk) \gamma_j$ as:
\be
&&h_0=m-t_0 \sum_{i=x,y,z} \cos k_i,\nonumber\\ 
&&h_{1,2,3}=t_{so}\sin( \sum_{j \neq i}k_j-k_i),\\
&&h_{4}=t_{so}\sin \sum_{i=x,y,z} k_i.\nonumber
\ee

\begin{figure}[htbp] 
	\centering
	\epsfig{file=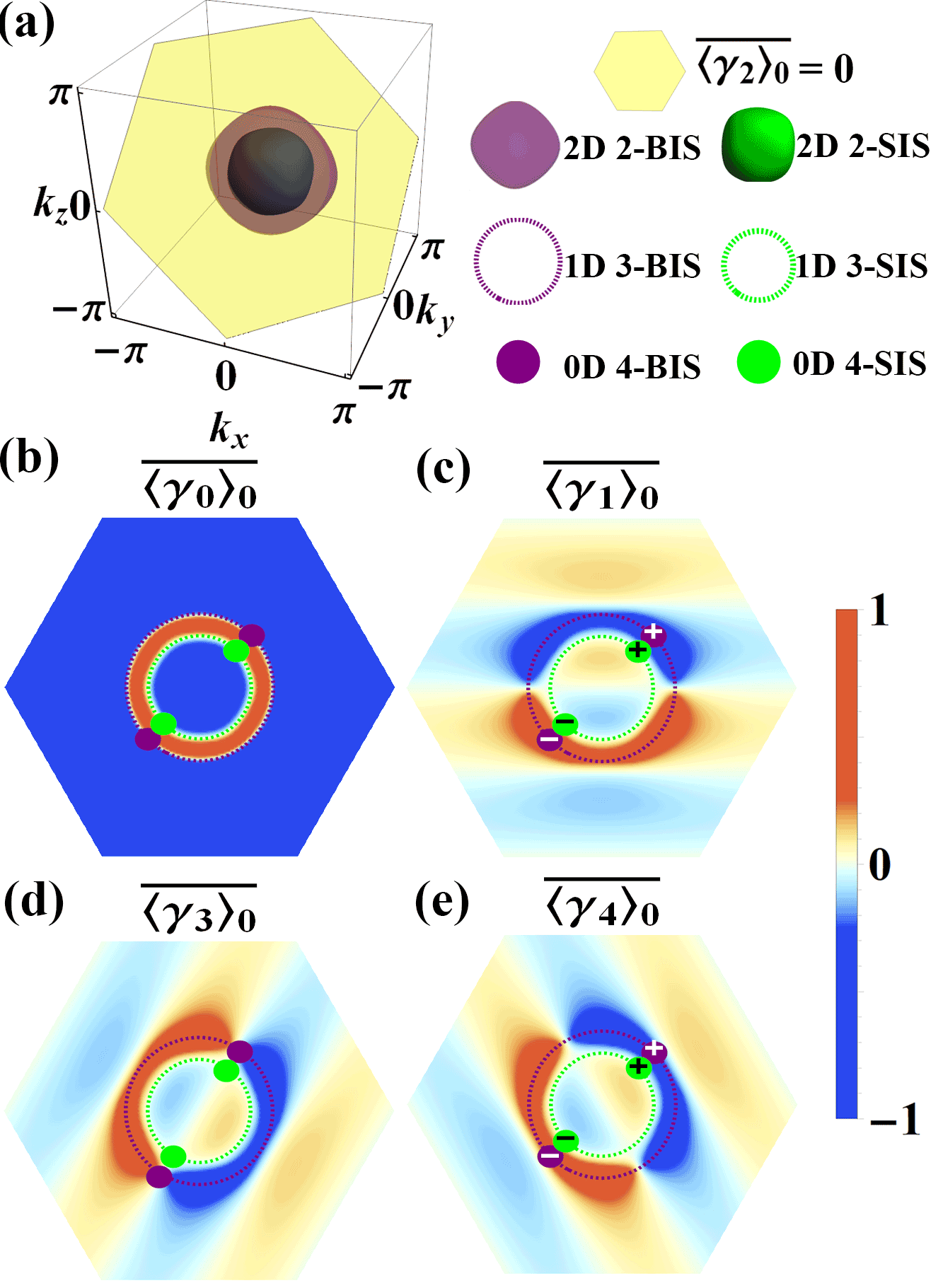, width=3.1in}
	\caption{Analytical  results of TASP for the 3D TR-invariant topological insulator in class AII with nonlinear quench protocol. The system is quenched from $t=0$ to $t=\infty$ with $5t_{so}=m/1.8=t_{0}=g=1$. (a) After quenching $h_0$, the nonzero overlap between the plane $\overline{\la { \gamma}_{2}\ra}_0=0$ and $\overline{\la { \gamma}_{0}\ra}_0=0$ are shown. (b) The corresponding TASP $\overline{\la { \gamma}_{0,1,3,4}\ra}_0$ on the plane $\overline{\la { \gamma}_{2}\ra}_0=0$. 3-SIS and  3-BIS are denoted by purple dashed lines and  green dashed lines, respectively. The opposite signs on 4-BIS (4-SIS) in $\overline{\la { \gamma}_{1,4}\ra}_0$  characterizes the nontrivial topology, and verifies the $\mathbb{Z}_2$ invariant $\nu^{(1)}$=$-1$.}
	\label{fig:3D_Z2}
\end{figure}
After only quenching $h_0$ axis, one can easily identify the 2-BIS and 2-SIS with vanishing spin polarization in $\overline{\la { \gamma}_{0}\ra}_0$, as the closed surfaces shown in Fig.~\ref{fig:3D_Z2} (a). Then 3-BIS and 3-SIS can be constructed by choosing the zero of $\overline{\la { \gamma}_{2}\ra}_0$ on 2-BIS and 2-SIS. Fig.~\ref{fig:3D_Z2} (b-e) show the other components of TASP $\overline{\la { \gamma}_{0,1,3,4}\ra}_0$ on the plane $\overline{\la { \gamma}_{2}\ra}_0=0$, and corresponding position of 3-BIS and 3-SIS are also indicated by the purple ring and green ring. Thus, the 0D 4-BIS (4-SIS) is constructed by the intersection of 3-BIS (3-SIS) and $\overline{\la { \gamma}_{3} \ra}_0=0$, on which the values (gradients) of the TASP in $\overline{\la { \gamma}_{1}\ra}_0$ and $\overline{\la { \gamma}_{4}\ra}_0$ all have opposite signs, verifying the nontrivial topological number $\nu^{(1)}$=$-1$.
\begin{figure}[htbp]
	\centering
	\epsfig{file=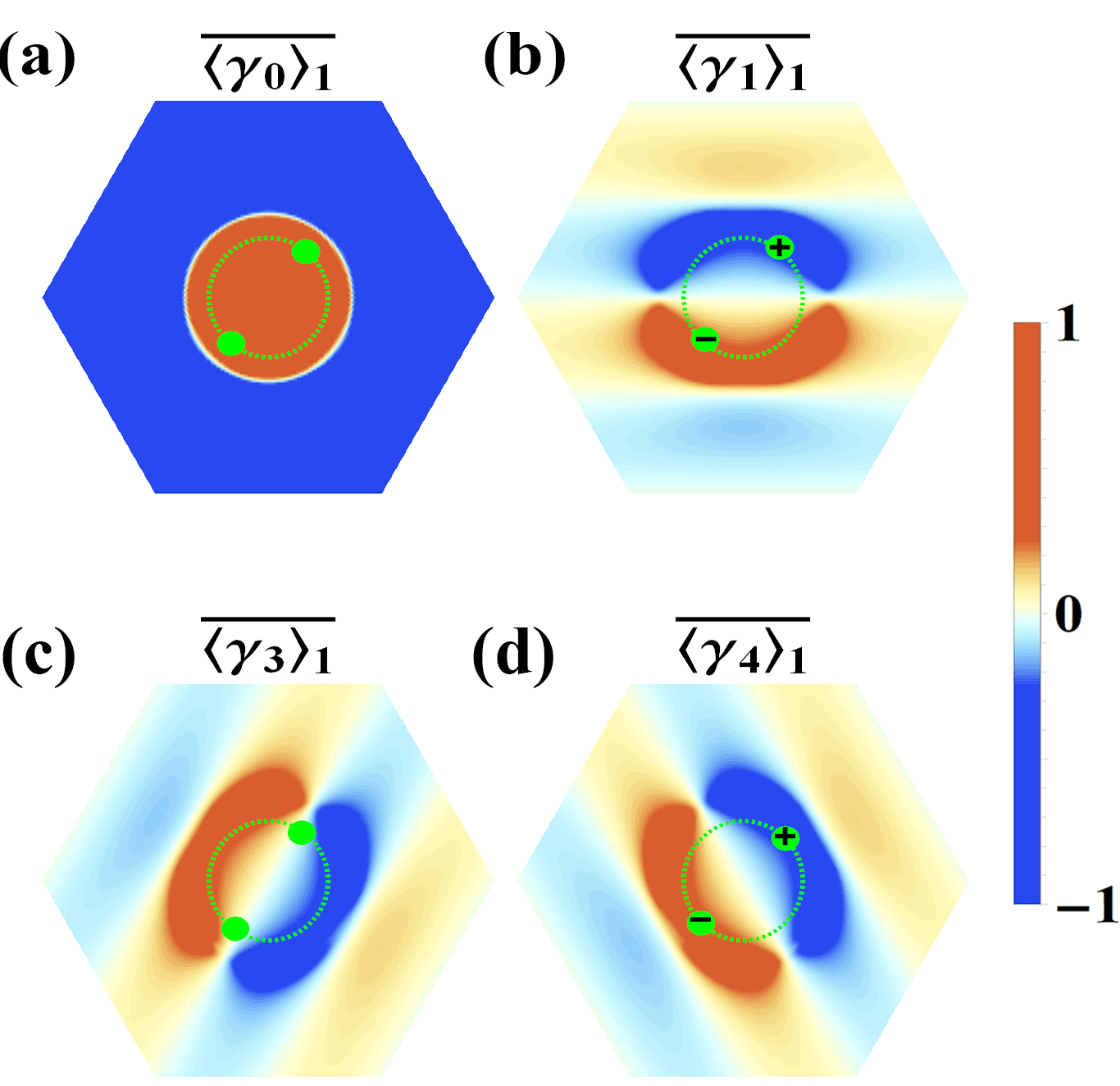, width=3.2in}
	\caption{The direct characterization based on SIS. The green line represents the 3-SIS of slowly quenching  $h_0$ axis. 0D 4-SIS is constructed by the 3-SIS and $\overline{\la {\gamma}_{3} \ra}_1=0$. The value of two spin textures $-\overline{\la {\gamma}_{1,4} \ra}_1$ on 4-SIS can form a opposite sign corresponding to nontrivial $\mathbb{Z}_2$ index $\nu^{(1)}$=$-1$.}
	\label{fig:SIS_tenfold_second}
\end{figure}

{\it The second scheme:} After determining the position of 2-SIS by deeply quenching the $h_0$ axis. We further perform a deep quench along another axis $h_1$, and obtain the corresponding TASP $\overline{\la { \gamma}_{0,1,3,4}\ra}_1$ on above plane $\overline{\la { \gamma}_{2}\ra}_0=0$. Then we can  construct the 4-SIS by the intersection of 3-SIS and $\overline{\la {\gamma}_{3} \ra}_1=0$. Thus, one can see a opposite sign given by the value $-\overline{\la {\gamma}_{1,4} \ra}_1$ on 4-SIS, implying a nontrivial $\mathbb{Z}_2$ index $\nu^{(1)}$=$-1$, as shown in Fig.~\ref{fig:SIS_tenfold_second}. 

\section{Discussion}
For comparison, we summarize the findings of the previous papers \cite{2018Sci,PRX2021,2021tenfold,2020us}  and this paper in Table.~\ref{table:comparision}. All the findings of this paper is indicated in red. The key points of this paper is : (i) Only in slow quench, there will be a SIS, (ii) The characterization  of topological phases based on SIS is twofold, (iii) The topological phases can be characterized by both SIS and BIS, and the direct characterization based on BIS and SIS can be realized.
\begin{table}[htbp]
	\centering
	\epsfig{file=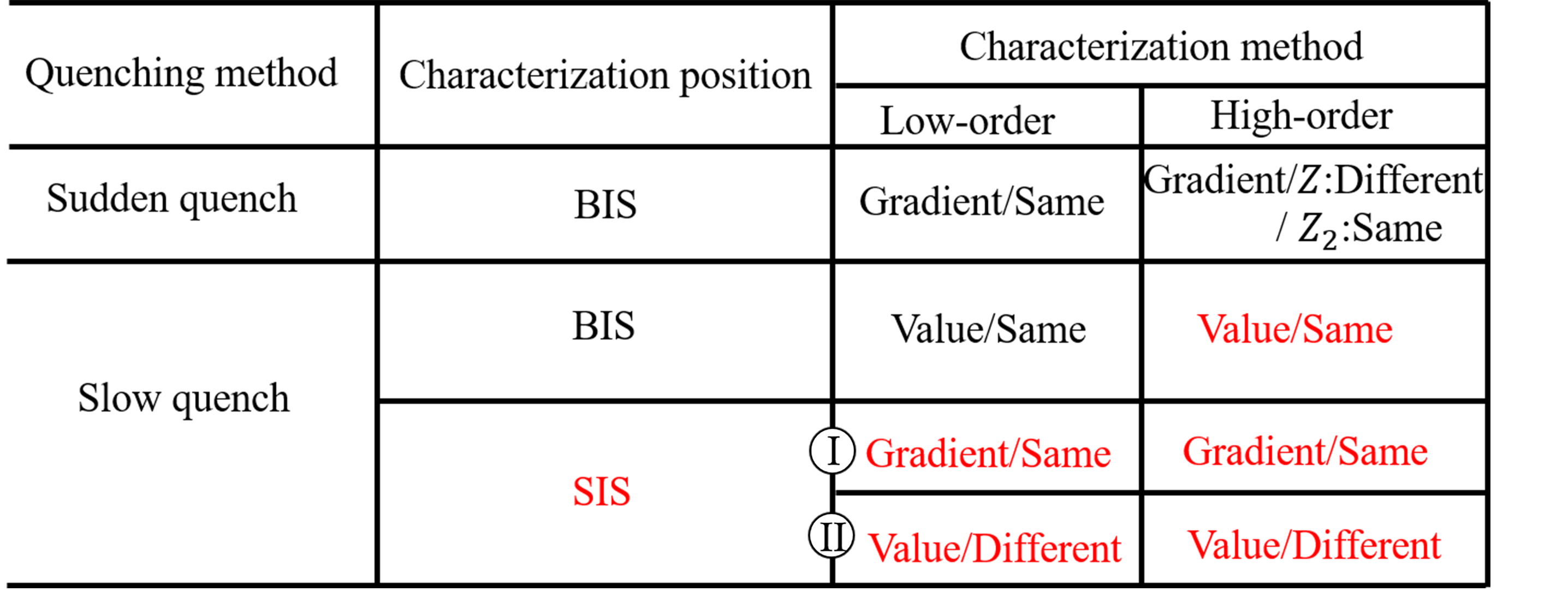, width=3.2in}
	\caption{The comparison of the previous study and the study in this paper. “Characterization method of lower-order” means the specific method of topological characterization  rely on low-order BIS or low-order SIS. “Characterization method of high-order” means the specific method of topological characterization  rely on  high-order BIS or  high-order SIS . In addition, “same”  mean the system is quenched once, and the topological characterization is in the same quenching process. “different”  means that two different quenching processes are needed to the topological characterization of the system.}
	\label{table:comparision}
\end{table}

\begin{figure}[htbp]
	\centering
	\epsfig{file=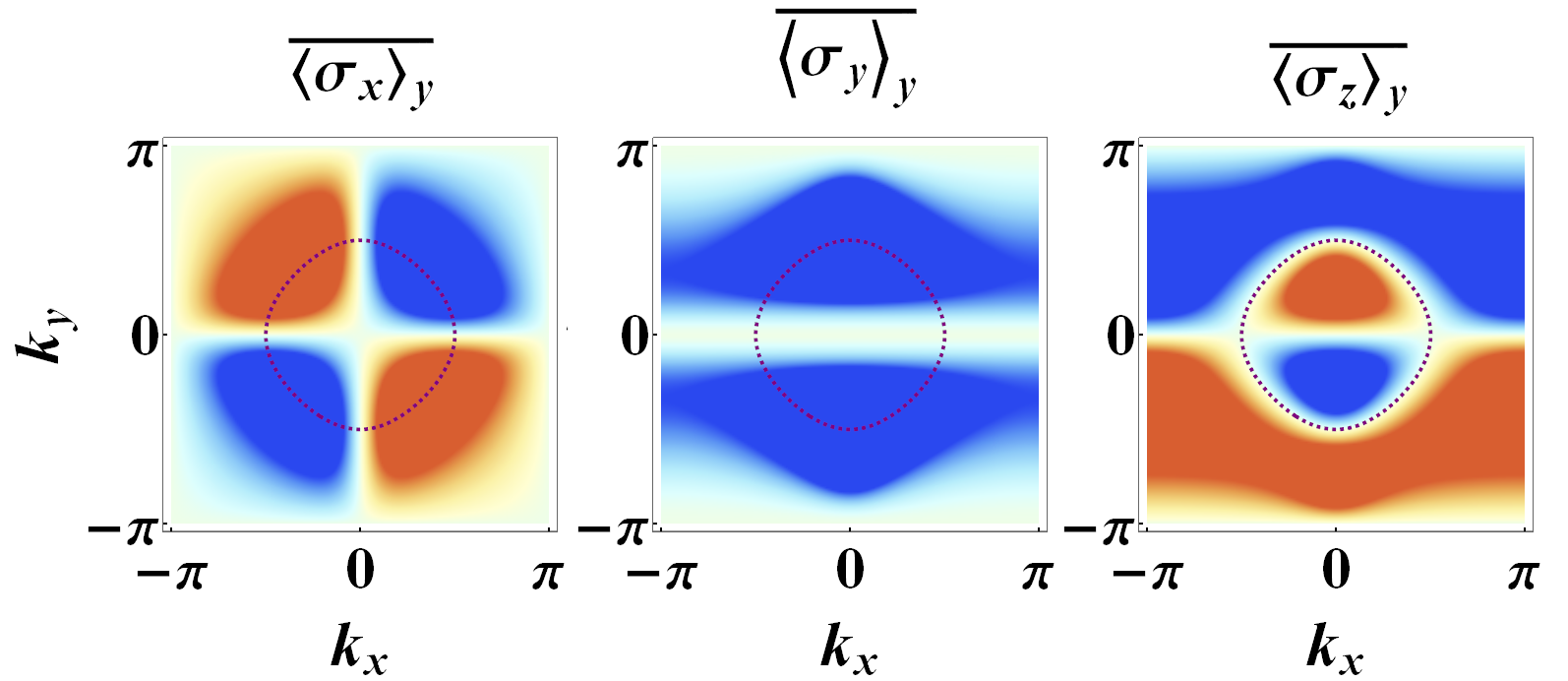, width=3.2in}
	\caption{Applying same procedures of the second scheme to the sudden quench. The purple line represents the BIS of sudden quench corresponding to $h_z=0$. $\overline{\la {\sigma}_{x,y,z} \ra}_y$ represent the TASP by suddenly quenching $m_y$ from $\infty$ to 0 with $m_z = t_0 =t_{so}^{x,y}=1$,$m_{x}=0$. The value of any two components of TASP cannot form a topological pattern on BIS.}
	\label{fig:BIS_second}
\end{figure}

It is noted that the above three traits are unique to the slow quench. To prove this, we apply same procedures of the second scheme to the sudden quench. Although the SIS in slow quench has similar characterization method (gradient) to BIS in sudden quench, one can obviously see that the value of TASP $\overline{\la {\sigma}_{x,y} \ra}_y$ on BIS is all less than or equal to zero, as shown in Fig.~{\ref{fig:BIS_second}}. Thus, the bulk topology cannot be  captured by the value of TASP on BIS.

{\section{Conclusions}}
In summary, we further unveil the topological characterization in the situation of slow nonadiabatic quench dynamics.  We approve both low-order SIS and BIS can capture the topological information of equilibrium state of system by only quenching one axis. Besides, a direct characterization of topological phases based on the low-order SIS is also presented by introducing another quenching process. Through a dimensional reduction approach, we apply our theory to high-order BIS and SIS that all give same topological information but flexible characterization method. Particularly, it is worthwhile to notice that the nonadiabatic quench is readily accessible in ultracold atomic experiments because the condition that the nonadiabatic transition time $t_c$ much smaller than the decoherence time of the ultracold atomic system $\tau$ can be satisfied. Thus, one can expect our characterization schemes may provide more efficient technique for future experiment.
\section*{Acknowledgements}
This work was supported by the National Key Research and Development Program of Ministry of Science and Technology (No. 2021YFA1200700), National Natural Science Foundation of China (No. 11905054, and No.11804122), the China Postdoctoral Science Foundation (Grant No. 2021M690970)  and  the Fundamental Research Funds for the Central Universities from China. 

\begin{appendix}	 
\section{Solving the multi-state LZ problem}
In this section, we provide the details in solving the multi-state LZ problem as given by the time-dependent Hamiltonian (\ref{eq:LZ}). Let's assume the state vector to be $|\Psi\ra = (c_1, c_2, c_3, c_4)^T$. The multi-state LZ problem can be explicitly written as: 
\be
i\frac{d}{dt} \left( \begin{array}{c}
c_1 \\
c_2 \\
c_3  \\
c_4
\end{array}
\right)
=  \left( \begin{array}{cc}
\frac{g}{t} + h_0 & \bh\cdot{\bm \sigma}   -i h_4\\
\bh\cdot{\bm \sigma}   + i h_4 & -\frac{g}{t} - h_0 
\end{array}
\right)   \left( \begin{array}{c}
c_1 \\
c_2 \\
c_3  \\
c_4
\end{array}
\right), \nn \\
\ee
with $\bh = (h_1, h_2, h_3)$.
By noting the identity $(\bh\cdot{\bm \sigma}   -i h_4) (\bh\cdot{\bm \sigma}  + i h_4))= h^2$ with $h^2 = \sum_{j=1}^4 h_j^2$, one can decouple the above four equations into two separated ones. First, one defines a new vector $(c_1', c_2')^T = \hat{u}(c_1, c_2)^T $ with $\hat{u} = (\bh\cdot{\bm \sigma}  + i h_4) /h $. Then one can obtain two coupled time-dependent problems in the basis of $(c_1', c_2')^T$ and $(c_3, c_4)^T$: 
\be
i\frac{d}{dt} \left( \begin{array}{c}
c_1' \\
c_2'
\end{array}
\right)
= \Big( \frac{g}{t} + h_0 \Big)   \left( \begin{array}{c}
c_1' \\
c_2'
\end{array}
\right) +  h  \left( \begin{array}{c}
c_3  \\
c_4
\end{array}
\right),
\ee
and 
\be
i\frac{d}{dt} \left( \begin{array}{c}
c_3 \\
c_4
\end{array}
\right)
= -\Big( \frac{g}{t} + h_0 \Big)   \left( \begin{array}{c}
c_3 \\
c_4
\end{array}
\right) +  h  \left( \begin{array}{c}
c_1'  \\
c_2'
\end{array}
\right). 
\ee
One can see that, now $c_1'$ and $c_3$ are coupled together, but they are decoupled from $c_2'$ and $c_4$. Thus we obtain two decoupled two-state LZ problem: 
\be
i\frac{d}{dt}  \left( \begin{array}{c}
a \\
b
\end{array}
\right)
=  \left( \begin{array}{cc}
\frac{g}{t} + h_0 & h\\
h   & -\frac{g}{t} - h_0 
\end{array}
\right)   \left( \begin{array}{c}
a  \\
b
\end{array}
\right),
\ee
whose solution has been obtained in our previous paper. Explicitly, starting from the initial ground state $(0, 1)^T$, the final state vector is written as: 
\be
|\psi(t) \ra = \sqrt{P}  e^{-i\ve t} |+ \ra +  \sqrt{1-P} e^{i\ve t + \phi_0} |-\ra
\ee
where $\phi_0$ is a relative phase factor that is independent of time. Here, $|\pm\ra$ are the two instantaneous eigenvectors of the two-level Hamiltonian $H = h_0 \sigma_z + h \sigma_x$, and $P$ is the transition probability: 
\be
P=\frac{e^{-2\pi g \cos\theta} - e^{-2\pi g}}{e^{2\pi g} - e^{-2\pi g}}
\ee
with $\cos\theta = h/\sqrt{h_0^2 + h^2}$. 
Returning to the basis of $(c_1, c_2, c_3, c_4)^T$, the solution (\ref{eq:psit}) can be readily obtained.

	\section{The definition of high-order BIS and high-order SIS}

The concept of high-order BIS was introduced with a dimension reduction approach induced by high-order DBSC \cite{PRX2021}. The first-order BIS (1${\raisebox{0mm}{-}}$BIS) is a $(d-1)$D momentum subspace with one of the components vanished in Hamiltonian. Then one can define $n${\raisebox{0mm}{-}}BIS on ($n-1$)-BIS by setting $h_{n-1}(\bk)=0$. In addition, the first-order SIS (1-SIS) is also a $(d-1)$D momentum subspace, on which $P_u-P_d=0$. The $n$-SIS is defined on ($n-1$)-SIS by also setting $h_{n-1}(\bk)=0$.  The highest-order BIS or SIS  consists of only several pairs of points, which greatly simplifies the characterization of topological phases. For convenience, we show the concept of high-order BIS and SIS  as follows:
\be
&&(d-n)D\quad n{\raisebox{0mm}{-}}BIS=\{h_0=h_1= \ldots h_{n-1}=0 \},\qquad \qquad
\ee
where $n\in\{1,d\}$.
\be
&&(d-n)D\quad n{\raisebox{0mm}{-}}SIS=\{P_u-P_d=h_1= \ldots h_{n-1}=0\},\qquad
\ee
where $n\in\{2,d\}$.

Analogous to the high-order SIS and BIS defined in $\mathbb{Z}$-type topological phases,  we can also give a similar definition in $\mathbb{Z}_2$ topological phases as follows:
	
for first descendant $\mathbb{Z}_2$ topological phases
\be
&&(d'-n)D\quad n{\raisebox{0mm}{-}}BIS=\{h_0=h_1= \ldots h_{n-2}=0 \},\qquad \qquad		
\ee
where $n\in\{2,d'=d+1\}$.
\be
&&(d'-n)D\quad n{\raisebox{0mm}{-}}SIS=\{P_u-P_d=h_1= \ldots h_{n-2}=0 \},\qquad
\ee
where $n\in\{3,d'=d+1\}$.

for second descendant $\mathbb{Z}_2$ topological phases
\be
&&(d'-n)D\quad n{\raisebox{0mm}{-}}BIS=\{h_0=h_1= \ldots h_{n-3}=0 \},\qquad \qquad
\ee	
where $n\in\{3,d'=d+2\}$.
\be
&&(d'-n)D\quad n{\raisebox{0mm}{-}}SIS=\{P_u-P_d=h_1= \ldots h_{n-3}=0 \},\qquad
\ee
where $n\in\{4,d'=d+1\}$.

The process of constructing high-order BIS and SIS in our main text coincides with definition here. According to the Eq.~(\ref{eq:spin polarization}), the component of TASP is $
\overline{\la { {\gamma}_i}\ra} = (P_u-P_d)n_i=(P_u-P_d)\frac{{h_i}} {\varepsilon}
$, thus the vanishing spin polarization in each component of TASP can be two cases: one is $h_i=0$, the other is $(P_u-P_d)=0$. In other words, each component not only has its unique spin-polarization $h_i=0$, but also the common spin-polarization $(P_u-P_d)=0$. Therefore, we define the $k$ points with  $(P_u-P_d)=0$ as the 1-SIS. Specifically, if we choose  $h_0$ as the quenching axis, thus in the $\overline{\la { {\gamma}_0}\ra}$, the region with $h_0=0$ corresponding to the 1-BIS and the region with $(P_u-P_d)=0$ corresponding to the 1-SIS. The intersection of  $\overline{\la { {\gamma}_0}\ra}=0$ and the zero in one of  
$\overline{\la { {\gamma}_i}\ra}$  actually can be divided into two cases: one is the intersection of $h_0=0$ and $h_i=0$, which constructs high-order BIS; the other is the intersection of $(P_u-P_d)=0$ and $h_i=0$, which constructs high-order SIS. The higher the order of BIS (SIS) is, the more components of TASP are needed for construction. When the dimension is reduced to 0D, the highest-order BIS (SIS) is constructed successfully.

\section{2D topological phases with high integer invariant}
\begin{figure}[htbp]
	\centering
	\epsfig{file=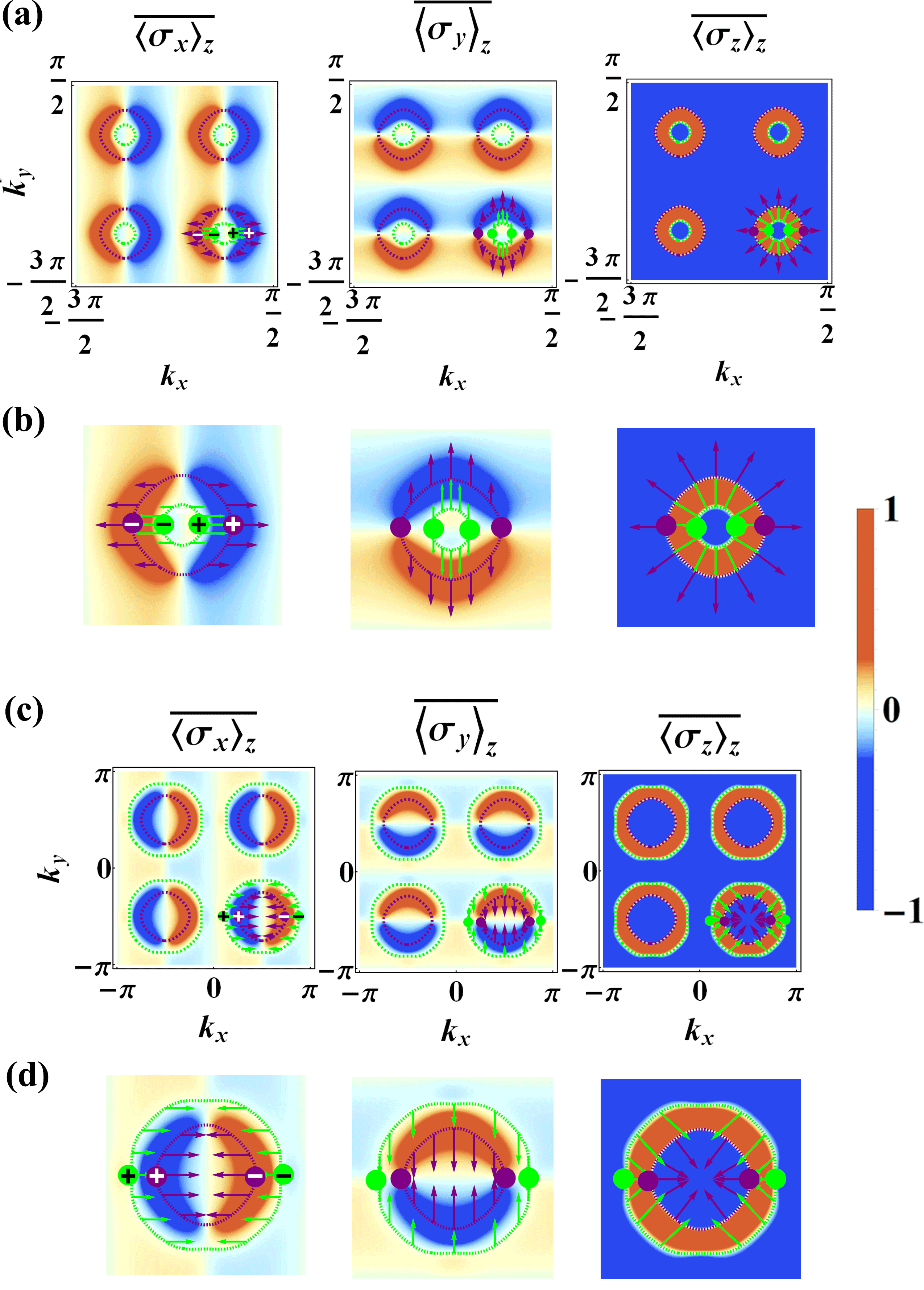, width=3.4in}
	\caption{Analytical results of TASP for the 2D topological model with high integer invariant. Three components $\overline{\la { \sigma}_{x}\ra}_z$ (left panel), $\overline{\la {\sigma}_{y}\ra}_z$ (middle panel) and $\overline{\la {\sigma}_{z}\ra}_z$ (right panel)  are shown, respectively. (a) The system is quenched from $t=0$ to $t\to\infty$ with $g=10$ and $5t_{so}=t_{0}=m_z=1$. (b) To get a better view, we give dynamical field in (a) separately. (c) The system is quenched from $t=0$ to $t\to\infty$ with $g=1$ and $5t_{so}=t_{0}=1,m_z=-1$. (d) To get a better view, we give dynamical field in (c) separately.}
	\label{fig:high_integer}
\end{figure}

To extend the validity of our findings, we apply the above characterization strategy to integer topological phases with invariant larger than $1$. Here, we consider a Hamiltonian with the variations of the 2D quantum anomalous Hall model: $\bm h(\bk)=(t_{so} \sin 2 k_x, t_{so} \sin 2k_y, m_z-t_0 \cos 2k_x-t_0 \cos 2k_y)$.  The topological phases of this model is determined by: $0<m_z<2t_0$ with the Chern number ${\cal C}_{1} = -4$, or $-2t_0<m_z<0$ with ${\cal C}_{1} = 4$.
For better description, we have shifted the Brillouin region to $(-3/2\pi,\pi/2)$ for $m_z=1$. As shown in Fig.~(\ref{fig:high_integer}) (a) and (b), the final TASP have four rings, and the dynamical field only wind once on each ring. Thus, the dynamical field of these four rings gives the Chern number ${\cal C}_{1} = -4$.
In Fig.~(\ref{fig:high_integer}) (c) and (d), we present the TASP and the corresponding dynamical field for $m_z=-1$. Compared with the dynamical field  shown in Fig .~(\ref{fig:high_integer}) (b), the dynamical field shown in Fig .~(\ref{fig:high_integer}) (d) converge at a central point inside BIS and SIS rather than spread out to the outside of BIS and SIS. Thus, the opposite Chern number ${\cal C}_{1} = 4 $ is given. In both cases, the 2-BIS and 2-SIS are constructed by the intersection of $\overline{\la {\sigma}_{y}\ra}_z=0$ and $\overline{\la {\sigma}_{z}\ra}_z=0$, on which the dynamical field has opposite signs, reflecting nontrivial topological number 4. The direct characterization scheme is similar as described in main text, and thus the results is not shown here.

\end{appendix}

\end{document}